\documentclass{elsart}

\begin{document}
\begin{frontmatter}

\title{Estimation of dibaryon
$(\Omega^{-}\Omega^{-})_{J^{\pi}=0^{+}}$ yields 
at RHIC energies} 
\author{Zhong-Dao~Lu}
\address{China Institute of Atomic Energy, P. O. Box 275(18), Beijing 102413, China, and}
\address{China Center of Advance Science and Technology (CCAST), Beijing 100080, China}

\begin{abstract}
The yields of dibaryon  $(\Omega\Omega)_{0^{+}}$ in relativistic 
heavy ion collisions at three energies (AGS, SPS and RHIC energies) 
are calculated in the framework of statistical model. 
It's yield at full RHIC energy, $\sqrt{s_{NN}}$ = 200 GeV,
is predicted by means of extrapolation. In the RHIC energy region 
(RHIC to full RHIC energies), 
it is in order of magnitude $10^{-4} - 10^{-3}$.
The yields of hyperon $\Omega^-$ and the ratios of  $(\Omega\Omega)_{0^{+}}$ to
$\Omega^-$ are also given. In the RHIC energy region, the yield 
of $\Omega^-$ is in order of $10^{0}-10^{1}$ and the ratio of 
$(\Omega\Omega)_{0^{+}}$ to $\Omega^-$ is $\simeq 1.0\times10^{-4}$.
\end{abstract}


\end{frontmatter}

\section{Introduction}

According to quantum chromodynamics (QCD) theory, there is the possibility
of the existence of some exotic multi-quark states, such as H dibaryon \cite{Jaffe77}, 
and gluonic states.
The growing interest of searching for such states lies on not only to examine 
the quantum chromodynamics (QCD) theory, but also to reveal the quark-gluon 
behavior in short distance and other new physical characteristics. 
But so far no convinced evidence is found in experiments to show their existence. 
Recently a new candidate, the strange dibaryon 
$(\Omega\Omega)_{0^{+}}$, is suggested and studied based on the 
chiral SU(3) quark model \cite{Zhang97,Shen97,Yu99,Zh00,Li01}. 
This six-quark states with large strangeness is found to have  
more attractive characteristics than
other six-quark states. First, it is a deeply bound state. Although the color magnetic 
interaction in the one gluon exchange term for this diomega system exhibits the 
repulsive feature, the large attraction stemming from the chiral gluon coupling 
and from the symmetry property of the system lead to a considerable binding. By 
the Resonating  Group Method, the binding energy of  $(\Omega\Omega)_{0^{+}}$ is 
found to be as large as $\approx$ 116 MeV, and the root-mean-square distance 
between two $\Omega^-$s is 0.84 fm. Second, $(\Omega\Omega)_{0^{+}}$ has a quite 
long lifetime of $\sim 10^{-10}$ sec as it can undergo only weak decay, such as 
$(\Omega\Omega)_{0^{+}} \rightarrow \pi^- + \Xi^0 + \Omega^-$ and 
$(\Omega\Omega)_{0^{+}} \rightarrow \pi^0 + \Xi^- + \Omega^-$. As they are three-body
decay, the final state phase-space would be suppressed. In the sudden approximation, 
the lifetime of $(\Omega\Omega)_{0^{+}}$ was found to be four times longer than 
the free $\Omega^-$ lifetime of 0.822$\times10^{-10}$ sec. Apart from these 
conventional decay modes, the non-meson 
decay $(\Omega\Omega)_{0^{+}} \rightarrow \Xi^0 + \Omega^-$ is also possible, the 
lifetime  of $(\Omega\Omega)_{0^{+}}$ for this process is twice the free $\Omega^-$ 
lifetime. Third, $(\Omega\Omega)_{0^{+}}$
has two negative charges. These properties make it 
easily be identified in experiment  when it were created.

Because of the large strangeness and very heavy mass, $(\Omega\Omega)_{0^{+}}$
is not likely to be produced in proton-proton collisions. While in relativistic
heavy ion collisions, it is found that the strangeness production is enhanced 
and has been suggested as one of the possible signals of quark-gluon-plasma 
(QGP). Recent experiments by the WA97 Collaboration \cite{AndC98} and the NA49 
Collaboration \cite{MargC99} showed the substantial enhancement of the (anti-)hyperon 
($\Lambda$, $\Xi$ and $\Omega$) yields in 158 A GeV/{c} Pb+Pb central collisions 
relative to p+Pb collisions, and exhibited the enhancement pattern increasing with
the strangeness content of the (anti-)hyperon. Therefore, the extreme environment
of very hot and very dense block of matter provided by the relativistic 
heavy ion collisions may lead to the formation of exotic deeply bound objects
composed of quarks or baryons with large strangeness, among them the dibaryon
$(\Omega\Omega)_{0^{+}}$ is the most possible candidate. 

Theoretically, the yields of  $(\Omega\Omega)_{0^{+}}$ 
can be estimated by two ways:
The kinetic simulation or microscopic transportation model, and 
the quantum statistical model or thermal model. Because the formation probability of 
$(\Omega\Omega)_{0^{+}}$ is very low 
and the cross sections related to $(\Omega\Omega)_{0^{+}}$
formations and decays are not well known, the calculations
by kinetic simulation can not be well carried out. Contrarily, 
the statistical model can do the calculations easier. 
Based on this model, ref.~5 gave out an estimation of 
$(\Omega\Omega)_{0^{+}}$ yields in Pb+Pb collisions 
at 158A GeV/$c$ (SPS), which is in order of magnitude $10^{-5}$. 
In this paper, we will
present its yields in other two reactions: Au+Au collision at 11.6A GeV/$c$
(AGS) and Au+Au collision at $\sqrt{s_{NN}}$ = 130 GeV (RHIC). And, based on the 
results of these three reactions, we try to predict the yield of 
 $(\Omega\Omega)_{0^{+}}$ for Au+Au collision
at full RHIC energy, $\sqrt{s_{NN}}$ = 200 GeV, by means of extrapolation. 
In order to reduce the effect of system size to the extrapolated results, the above  
collisions are chosen to 
have the same or similar reaction system.
The yields of $\Omega^-$ and 
the ratios of  $(\Omega\Omega)_{0^{+}}$ to $\Omega^-$ are also given.

\section{Thermal Production}

In relativistic heavy ion collisions, a large amount of particles 
including baryons, mesons, their resonance and anti-particles are 
produced and form a fireball of hadronic gases with high 
particle density and high energy density. The fireball
undergoes the expansion until freeze out. The large 
amount of produced particles 
and the violent collisions between them lead the gaseous particles 
inside the fireball to reach the chemical and thermal equilibrium. 
Such gas system 
can be described by the statistical model or thermal model. This model
has been successfully applied in analysis of hadron yields and ratios
at AGS and SPS energies 
\cite{BrSt95,ClRe01,YeGo99,BHS99,ClRe99,LeRa99,Lu98,Yen97,Lu02,LuRt01}, 
and recently the RHIC energy \cite{BrMrhic,Lu01}.
Here we use this model to estimate the yields of  
$(\Omega\Omega)_{0^+}$ in above three energies and predict it's yield in 
full RHIC energy, $\sqrt{s_{NN}}$ = 200 GeV, by means of extrapolation. 

In the frame of grand canonical ensemble, the density of hadron species i
reads as  (in unit of $c = \hbar = k_B = 1$)  
\begin{equation}
\rho_i = \frac{g_i}{(2\pi)^3}\int_0^{\infty}\frac{ d^3q}
{e^{(\varepsilon-\mu_i)/T}\pm 1}\ ,\\
\label{dens1}
\end{equation}
where T is the temperature of the fireball,
$\varepsilon=\sqrt{q^2+m_{i}^{2}}$ is the energy of the particle,
q and $m_i$ are the momentum and mass, $g_i$ is the spin-isospin degeneracy.
$\pm$ stands for the characteristics of statistics, + is for the Fermions and 
- is for the Bosons. 
The chemical potential is determined by $\mu_i = B_{i}\mu_{B} + S_{i}\mu_{S}$, 
$\mu_{B}$ and $\mu_{S}$ are the chemical potentials of baryon and 
strangeness respectively, $B_{i}$ and $S_{i}$ are the baryon charges and 
strangeness charges of the particle respectively.

Since the temperature of the fireball
is very high, it is reasonable to treat the hadron gas as
the ideal gas obeying the Boltzmann distribution. In the condition of 
$m/T \gg 1$, the density of species i can be well approximated by
\begin{equation}
\rho_i=\,g_i\,(\frac{m_{i} T}{2\pi})^{3/2} \, e^{-(m_i-\mu_i)/T}\ .  \\  
\label{dens2}
\end{equation}

We carried out the calculations for the following three relativistic heavy ion collisions: 
Au+Au collision at 11.6A GeV/$c$ (AGS), 
Pb+Pb collision at 158A GeV/$c$ (SPS) and Au+Au collision 
at $\sqrt{s_{NN}}$ = 130 GeV (RHIC). 
The experimental data of the hadron yields and ratios are adopted from 
\cite{ClRe01,BHS99,BrMrhic} and the references therein, respectively. 
All hadrons including baryons, mesons and 
possible anti-particles with mass less than 2 GeV/$c^2$ are included in the 
hadron gas of the fireball. The calculations are carried out within the frame of 
two-source statistical model (TSM) \cite{Lu02,LuRt01,Lu01}. 
This model is developed on the basis of 
the standard statistical model, the single-source model (SM), and takes 
the non-homogeneous effect of the source into account. 
This effect is showed up in 
the experimental observation of low baryon components in the
midrapidity region of relativistic heavy-ion collisions at energy
of 158A GeV/$c$ (see, e.g. \cite{na52}) and 
in the microscopic model calculations 
which show the central zone is more baryon dilute than outside zones 
\cite{Brprc99,SHSX99}. 
This effect of "non-homogenous" implies the requirement of a 
multiple source model to better 
describe the source
of a hadronic gas. The two-source model (TSM)
is the simplest multiple source model.  
In TSM the 
whole source is roughly divided into two regions (sources): the inner source $S_{2}$ 
and the outer source $S_{1}$. The two sources are supposed to reach equilibrium 
respectively and possess different temperatures, densities, chemical potentials and 
other thermodynamic characteristics. Especially the net strangeness density 
is no longer zero everywhere as in the single-source model but the total number
of strangeness still keeps conserved. 
The two-source model significantly improves the agreement to the experimental 
data of hadron yields and ratios and exhibits a reasonable source structure 
\cite{Lu02,LuRt01,Lu01}: 
At SPS and AGS energies, the whole source is composed of a small, 
hotter core (inner source) surrounded by a large, cooler halo (outer source). 
Most baryons are distributed outside, seems to be the projectile-
and target-like components, while almost all anti-baryons are 
concentrated inside.
Besides, in the two-source model approach, the average particle densities and 
the average energy density are effectively reduced and the strangeness
is effectively suppressed but without introducing extra corrections, such as the
corrections of hard core excluded volume of the particles and the strangeness suppression.
These corrections were generally used when the whole source is assumed to 
be a unique homogeneous source.

By means of least square method and by fitting the experimental data, 
the main thermodynamic characteristics in the two sources, S2/S1, are obtained as follows:

{\it AGS energy} \, the temperatures, T2/T1, are 144/91 MeV. The baryon 
chemical potentials, 
$\mu_{B2}/\mu_{B1}$, are 288/599 MeV. The strangeness chemical potentials, 
$\mu_{S2}/\mu_{S1}$, are -70.0/116 MeV. 

{\it SPS energy} \, T2/T1=168/110 MeV, 
$\mu_{B2}$/$\mu_{B1}$=14.4/387 MeV, 
$\mu_{S2}$/$\mu_{S1}$=-14.8/52.2 MeV. 

{\it RHIC energy} \, T2/T1=176/175 MeV, 
$\mu_{B2}$/$\mu_{B1}$=36.9/38.9 MeV, 
$\mu_{S2}$/$\mu_{S1}$=10.0/10.7 MeV. 

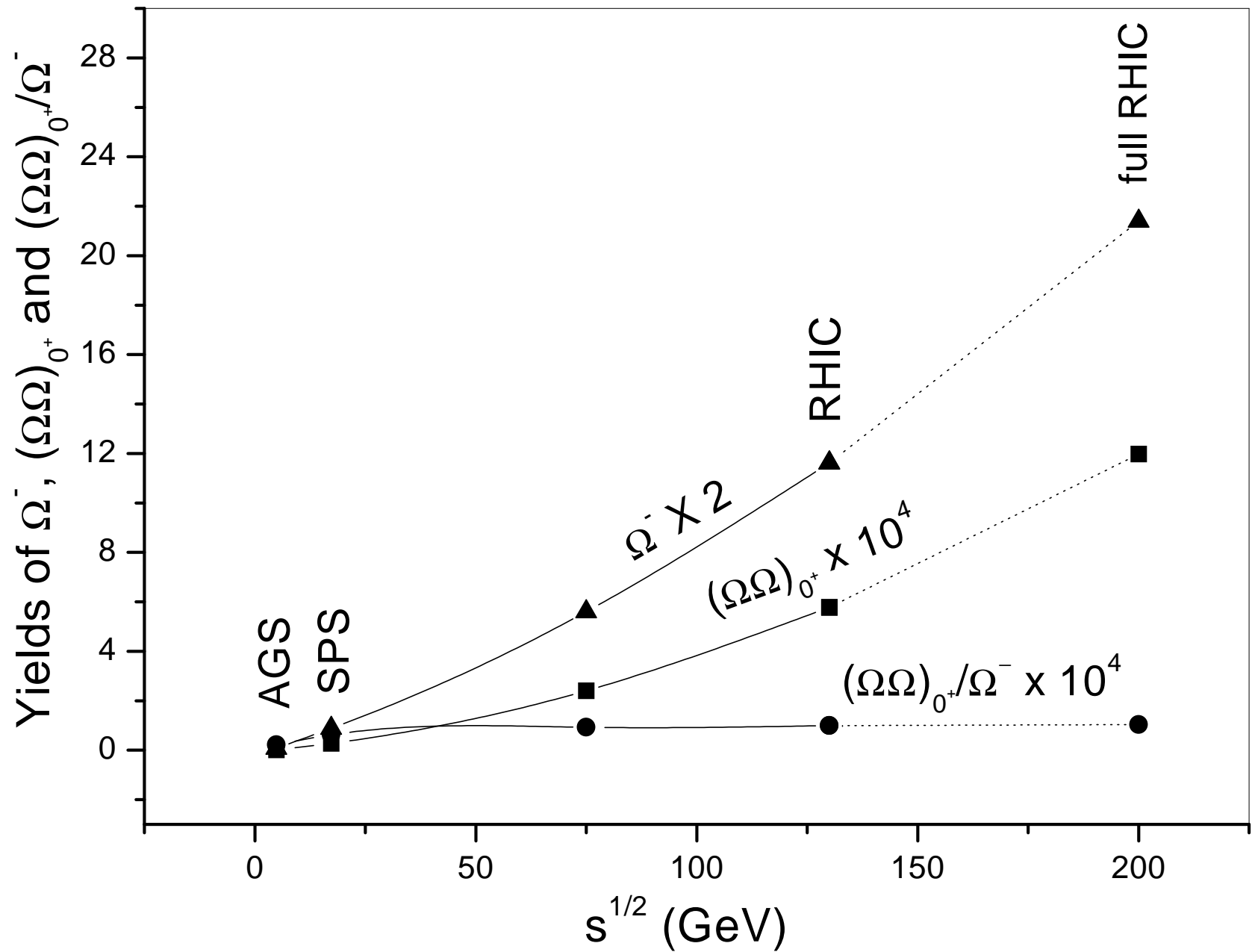
\begin{figure}
\vspace{50mm}
\caption{
Energy dependence of $(\Omega\Omega)_{0^+}$,
$\Omega^-$ and $(\Omega\Omega)_{0^+}/\Omega^-$ 
in relativistic heavy ion collisions. 
Points at full RHIC energy, $\sqrt{s_{NN}}$=200 GeV, 
are obtained by extrapolation. Points at $\sqrt{s_{NN}}$=75 GeV are obtained 
by interpolation. 
}
\label{fig1}
\end{figure}

The $\Omega^-$ yields in above three reactions 
are directly obtained through formula~\ref{dens2}. 
Because of the large strangeness, -6, and very heavy mass, 3230 MeV, the 
$(\Omega\Omega)_{0^+}$ yields are very small.
Therefore, it can be supposed that 
the thermodynamic quantities would be little effected 
by the existence of $(\Omega\Omega)_{0^+}$. In statistical model of 
grand canonical frame, all hadrons in the fireball including 
$(\Omega^{-}\Omega^{-})_{0^+}$ and $\Omega^-$ are in chemical and 
thermal equilibrium. 
According to formula~\ref{dens2}, the yield ratio of 
$(\Omega\Omega)_{0^+}$ to $\Omega^-$ has the relation of
\begin{equation}
\eta=\,\frac{g_{(\Omega\Omega)}}{g_{\Omega}}\,
(\frac{m_{(\Omega\Omega)}}{m_{\Omega}})^{3/2} \, 
e^{-[(m_{(\Omega\Omega)} -\mu_{(\Omega\Omega)}) - (m_{\Omega} -\mu_{\Omega})]} \ . \\
\label{ratio}
\end{equation}
With formulas~\ref{dens2} and~\ref{ratio} the $(\Omega\Omega)_{0^+}$  yields  
can be obtained from 
the  $\Omega^-$ yields.
The yields of $(\Omega\Omega)_{0^+}$ and $\Omega^-$ 
together with the ratios are presented in Table~\ref{tab1} (rows 2-4) and 
shown in Fig.~\ref{fig1}.

\vspace{10mm}
\begin{center} (put Table~\ref{tab1} here) \end{center}
\vspace{10mm}

Apparently, the yields of $(\Omega\Omega)_{0^+}$ and $\Omega^-$ increase steadily
as the energy increases from AGS to SPS, and to RHIC. At RHIC energy, they are 
5.8$\times10^{-4}$ and 5.8 respectively. 
The ratio, however, shows 
a saturation feature in RHIC energy region 
with value of $\simeq 1.0\times10^{-4}$. 

With results in above three reactions, we can try to predict the results 
for Au+Au collisions at 
full RHIC energy, $\sqrt{s_{NN}}$ = 200 GeV, by means of extrapolation 
(with Lagrange formula). 
In order to reduce the effect of system size to the extrapolated results as possible, 
these three collisions have been chosen to possess the same or 
similar reaction system. 
The results are shown in Fig. 1 (last points) and also shown in Table 1 (last column). 
At RHIC and full RHIC energies, the yields of $(\Omega\Omega)_{0^+}$ 
are 5.8$\times10^{-4}$ and 1.2$\times10^{-3}$ respectively, well within 
the limits of the detectors used in RHIC energies. With the excellent characteristics, 
$(\Omega\Omega)_{0^+}$, the marvelous particle, can be easily identified and 
detected when it is created. 

\section{Discussion and Summary}

In order to justify the reliability of the model predictions, we give out 
some results related to the strange particles and compare them to the experimental 
data or other model predictions. Table~\ref{tab2} and Fig.2 show the yields of 
$\overline{\Omega}^+$, $\Omega^-$ and their ratio 
$\overline{\Omega}^{+}/\Omega^-$. At low energy the $\overline{\Omega}^+$ yield is 
much lower than the $\Omega^-$ yield, while $\overline{\Omega}^+$  increases 
faster than $\Omega^-$ as the energy increases. So at somewhere they will have 
equal yield (in the figure, it is at $\sqrt{s_{NN}}\sim$ 150 GeV). 
 Table~\ref{tab3} and Fig.3 show the ratios of the anti-strange particle to the 
strange particle. In Fig.3 the results at AGS energy are for 11.6A GeV/$c$ Au+Au 
collision used for extrapolations. While in Table~\ref{tab3} 
the results at AGS energy are 
for 14.6A GeV/$c$ Si+Au collision used to compare with the available 
experimental data\cite{BrSt95}. For the strange baryons, the ratios possess 
the similar behavior 
as the energy increases but the magnitudes are different. Ratio for particle with 
more strange components is larger than that with less strange components. The 
ratio of kaon exhibits a different feature which saturates 
after $\sqrt{s_{NN}}$ = 70 GeV with 
value of 0.8. From the comparisons in Tables~\ref{tab2} and~\ref{tab3}, 
one can see that the model predictions are 
in very good agreement with the experimental data.

\vspace{10mm}
\begin{center} (put Table~\ref{tab2} here) \end{center}
\vspace{10mm}
\begin{center} (put Table~\ref{tab3} here) \end{center}
\vspace{10mm}

\begin{figure*}

\begin{center}(put two figures in the same row) \end{center}

\vspace{50mm}
\caption{
Energy dependence of $\overline{\Omega}^{+}$,
$\Omega^-$ and $\overline{\Omega}^{+}/\Omega^-$
in relativistic heavy ion collisions. 
Points at full RHIC energy 
are obtained by extrapolation. Points at $\sqrt{s_{NN}}$=75 GeV are obtained 
by interpolation.
} 
\label{fig2}
\caption{
Energy dependence of $\overline{\Lambda}/\Lambda$,
$\overline{\Xi}^{+}/\Xi^-$, $\overline{\Omega}^{+}/\Omega^-$ and $K^{-}/K^{+}$
in relativistic heavy ion collisions. 
Points at full RHIC energy 
are obtained by extrapolation. Points at $\sqrt{s_{NN}}$=75 GeV are obtained 
by interpolation. 
}
\label{fig3}
\end{figure*}

Several points need be pointed out. 
Firstly, the yields and ratios obtained above are thermally produced, 
i.e. produced in a hot and dense 
fireball which is assumed to reach the equilibrium or local equilibrium. 
If taking the kinetic effect into account, some amount
of $(\Omega\Omega)_{0^+}$
would be produced from two $\Omega^{-}$s' collision, the total
yield of $(\Omega\Omega)_{0^+}$ will be a little increased. 
Secondly, the yields of $\Omega^{-}$ predicted by 
statistical models generally lower than that given by experiments. So
the yields of $\Omega^{-}$ and $(\Omega\Omega)_{0^+}$ presented here 
are the lower limit.
Thirdly, more importantly, the experimental data used in calculations 
for RHIC energy are 
taken from pseudorapidity interval $|\eta|<1$ which is a small portion of 
the whole source of which the rapidity range is $-5.4<y<5.4$. 
(This small portion of source, the "measured" source, is located in the central part 
of the source. In this region, the particles are uniformly distributed 
and construct a single homogeneous source. So the two parts, S1 and S2, in TSM are
identical (see above the thermodynamic characteristics at RHIC energy) and also 
identical to the source in the single-source model.)
Therefore, if the data are taken from the whole source or a larger 
rapidity interval, the
yields of $\Omega^{-}$ will be raised and correspondingly,  
the yields of $(\Omega\Omega)_{0^+}$ increases. For example, 
in a conservative estimation, 
suppose the yield of $\Omega^{-}$ increases two times to reach 12, 
the yield of $(\Omega\Omega)_{0^+}$ increases the same times to
reach 1.2$\times10^{-3}$. At full RHIC energy, their yields increase $\sim2.2$
times and reach 24 and 2.6$\times10^{-3}$ respectively.
Obviously, a larger central rapidity range is favorable to measuring the 
interesting but 
rarely produced strange dibaryon, $(\Omega\Omega)_{0^+}$. 
Of cause, the more exact results are expected after the data of full RHIC 
energy become available. 

So far experimental data are available only for three collisions at AGS, 
SPS and RHIC energies with the same or similar collision system. There is 
no data available within the large energy region between SPS and RHIC 
energies. The particle yields, not as the ratios, are quite sensitive to the
"measured" source size or "measured" rapidity interval. Under these limitations, the
yields are only roughly estimated and may have large deviation, but it is believable 
that the order of magnitude would not deviate far away. 

In summary,
the yields of dibaryon $(\Omega\Omega)_{0^{+}}$, hyperon $\Omega^-$ 
and their ratios in relativistic 
heavy ion collisions at three energies are calculated within 
the framework of two-source statistical model. 
The yields of $(\Omega\Omega)_{0^{+}}$ and $\Omega^-$ increase steadily 
as the energy increases, while their ratios display a saturation feature. 
By means of extrapolation, the yields and the ratio at 
full RHIC energy are estimated. In the RHIC energy region, 
the  $(\Omega\Omega)_{0^{+}}$ yield is 
in order of $10^{-4} - 10^{-3}$, the 
$\Omega^-$ yield is in order of $10^{0}-10^{1}$ and the ratio of 
$(\Omega\Omega)_{0^{+}}$ to $\Omega^-$ is $\simeq 1.0\times10^{-4}$.
And, to measure the rarely produced particles as the 
strange dibaryon $(\Omega\Omega)_{0^+}$,
a larger central rapidity range is favorable.

{\bf Acknowledgements.}
The author are grateful to Professors Z.Y. Zhang,Y.W. Yu, C.R. Qing and H.Z. Xo for 
stimulating discussions. 
This work is supported by the National Science Foundation of China under the contracts
No.19975075.

\newpage

\begin{center}
FIGURE CAPTIONS
\end{center}

  Fig 1 Energy dependence of $(\Omega\Omega)_{0^+}$,
$\Omega^-$ and $(\Omega\Omega)_{0^+}/\Omega^-$ 
in relativistic heavy ion collisions. 
Points at full RHIC energy, $\sqrt{s_{NN}}$=200 GeV, 
are obtained by extrapolation. Points at $\sqrt{s_{NN}}$=75 GeV are obtained 
by interpolation. 

  Fig 2 Energy dependence of $\overline{\Omega}^{+}$,
$\Omega^-$ and $\overline{\Omega}^{+}/\Omega^-$
in relativistic heavy ion collisions. 
Points at full RHIC energy 
are obtained by extrapolation. Points at $\sqrt{s_{NN}}$=75 GeV are obtained 
by interpolation. 

  Fig 3 Energy dependence of $\overline{\Lambda}/\Lambda$,
$\overline{\Xi}^{+}/\Xi^-$, $\overline{\Omega}^{+}/\Omega^-$ and $K^{-}/K^{+}$
in relativistic heavy ion collisions. 
Points at full RHIC energy 
are obtained by extrapolation. Points at $\sqrt{s_{NN}}$=75 GeV are obtained 
by interpolation.

\newpage

\begin{table}
\caption{
Yields of $(\Omega^{-}\Omega^{-})_{0^+}$ and $\Omega^-$, and their
 ratios in relativistic heavy ion collisions. 
Results at full RHIC energy, $\sqrt{s_{NN}}$=200 GeV, are obtained 
by extrapolation. 
}
\vspace*{1ex}
\begin{tabular}{|c|c|c|c|c|}     \hline\hline
     & AGS & SPS & RHIC & full RHIC \\ \hline
$(\Omega^{-}\Omega^{-})_{0^+}$ & 9.18$\times10^{-7}$ & 2.78$\times10^{-5}$ & 
  5.78$\times10^{-4}$ & 1.20$\times10^{-3}$ \\ 
$\Omega^-$ & 0.0423 & 0.444 & 5.81 & 10.7 \\
$\eta$ & 2.17$\times10^{-5}$ & 6.26$\times10^{-5}$ & 
  0.996$\times10^{-4}$ & 1.03$\times10^{-4}$ \\ \hline\hline
\end{tabular}
\label{tab1}
\end{table}

\vspace{1.5cm}

\begin{table}
\caption{
Yields of $\Omega^{-}$ and $\overline{\Omega}^{+}$, 
and $\overline{\Omega}^{+}/\Omega^-$ in relativistic heavy ion collisions. 
Results at full RHIC energy are obtained by extrapolation. 
}
\vspace*{1ex}
\begin{tabular}{|c|c|c|c|c|}     \hline\hline
     & AGS & SPS & RHIC & full RHIC \\ \hline
$\Omega^-$ & 4.23$\times10^{-2}$ & 0.444 & 5.81 & 10.7 \\
$\overline{\Omega}^{+}$ & 3.74$\times10^{-3}$ & 0.166 & 
  5.37 & 12.0 \\ 
$\overline{\Omega}^{+}/\Omega^-$ & 8.84$\times10^{-2}$ & 0.378 & 
0.926 & 1.04 \\ 
&     &  exp: 0.383$\pm$0.081\cite{Ander99,BHS99}  & (0.898\cite{BrMrhic})
 &  (0.941\cite{BrMrhic}) \\ \hline\hline
\end{tabular}
\label{tab2}
\end{table}

\vspace{1.5cm}

\begin{table}
\caption{
Ratios of anti-strange particle to strange particle 
in relativistic heavy ion collisions. 
Results at full RHIC energy are obtained by extrapolation. 
}
\vspace*{1ex}
\begin{tabular}{|c|c|c|c|c|}     \hline\hline
     & AGS$^*$ & SPS & RHIC & full RHIC \\ \hline
$\overline{\Lambda}/\Lambda$ & 1.40$\times10^{-3}$  & 0.134 & 0.738 & 0.930 \\
     & $(2.0\pm0.8)\times10^{-3}$ \cite{Steph94,BrSt95}  
& 0.131$\pm$0.017\cite{Ander99,BHS99} 
& 0.77$\pm$0.07\cite{ZXu01,BrMrhic} &        \\
$\overline{\Xi}^{+}/\Xi^{-}$ & 1.74$\times10^{-3}$ & 0.211 & 
  0.823 & 1.00 \\ 
     &  & 0.232$\pm$0.033\cite{Appel98,BHS99} & 0.82$\pm$0.08\cite{BrMrhic} &   \\
$\overline{\Omega}^{+}/\Omega^-$ & 8.12$\times10^{-4}$ & 0.378 & 
0.926 & 1.04 \\ 
&     & 0.383$\pm$0.081\cite{Ander99,BHS99} & (0.898\cite{BrMrhic})
 &  (0.941\cite{BrMrhic}) \\ 
$K^{-}/K^{+}$ & 4.54$^{-1}$ & 1.82$^{-1}$ & 
  0.800 & 0.791 \\ 
 & $(4.4\pm0.4)^{-1}$\cite{Abb94,BrSt95} & $(1.8\pm0.1)^{-1}$\cite{Ander99,BHS99}
 & 0.88$\pm$0.07$^{**}$  & \\ 
 \hline\hline
\end{tabular}
\label{tab3}

* Both data and results are for Si+Au collision at 14.6A GeV/{c}. \\
** Average value of four experimental data shown in \cite{BrMrhic}. 
\end{table}

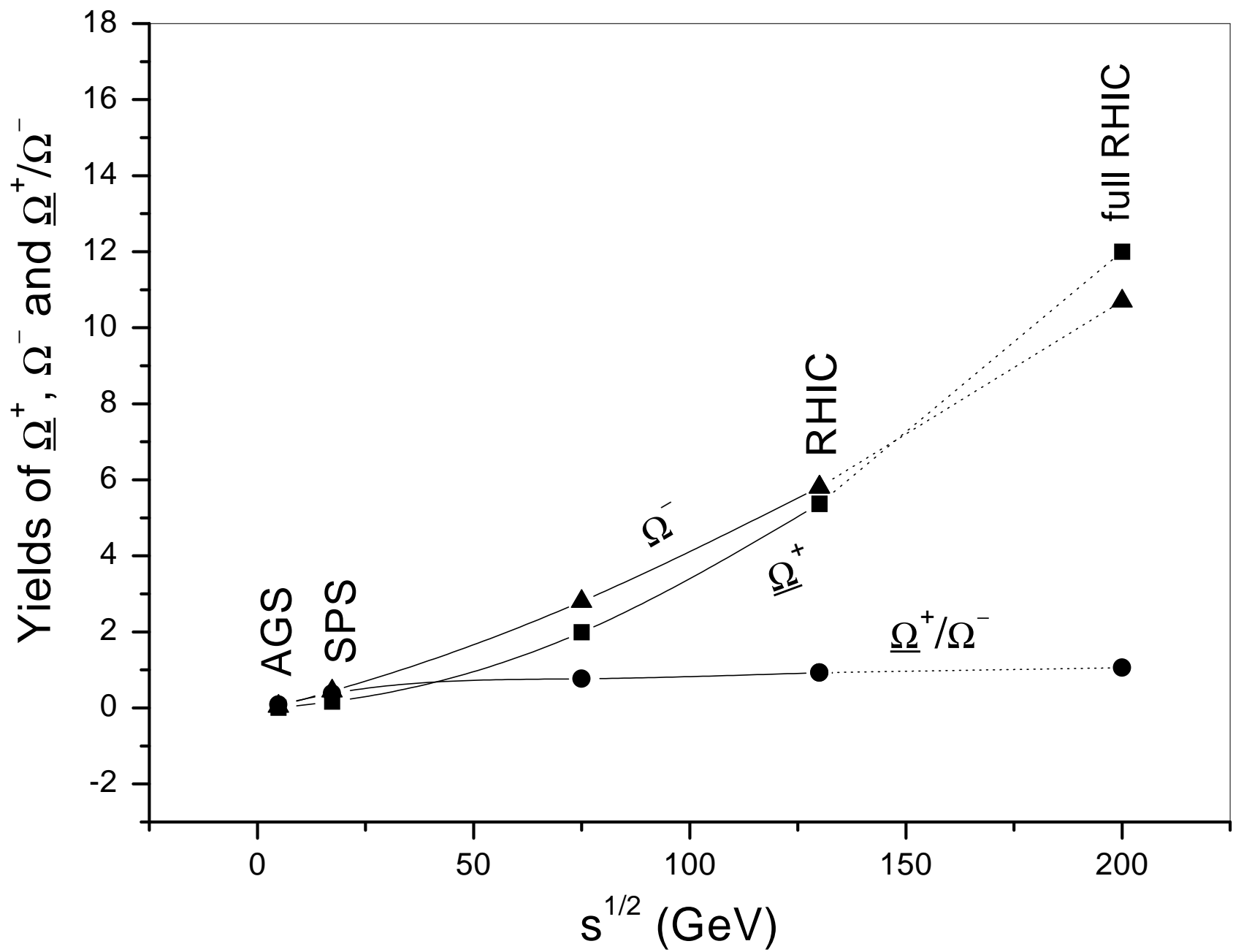
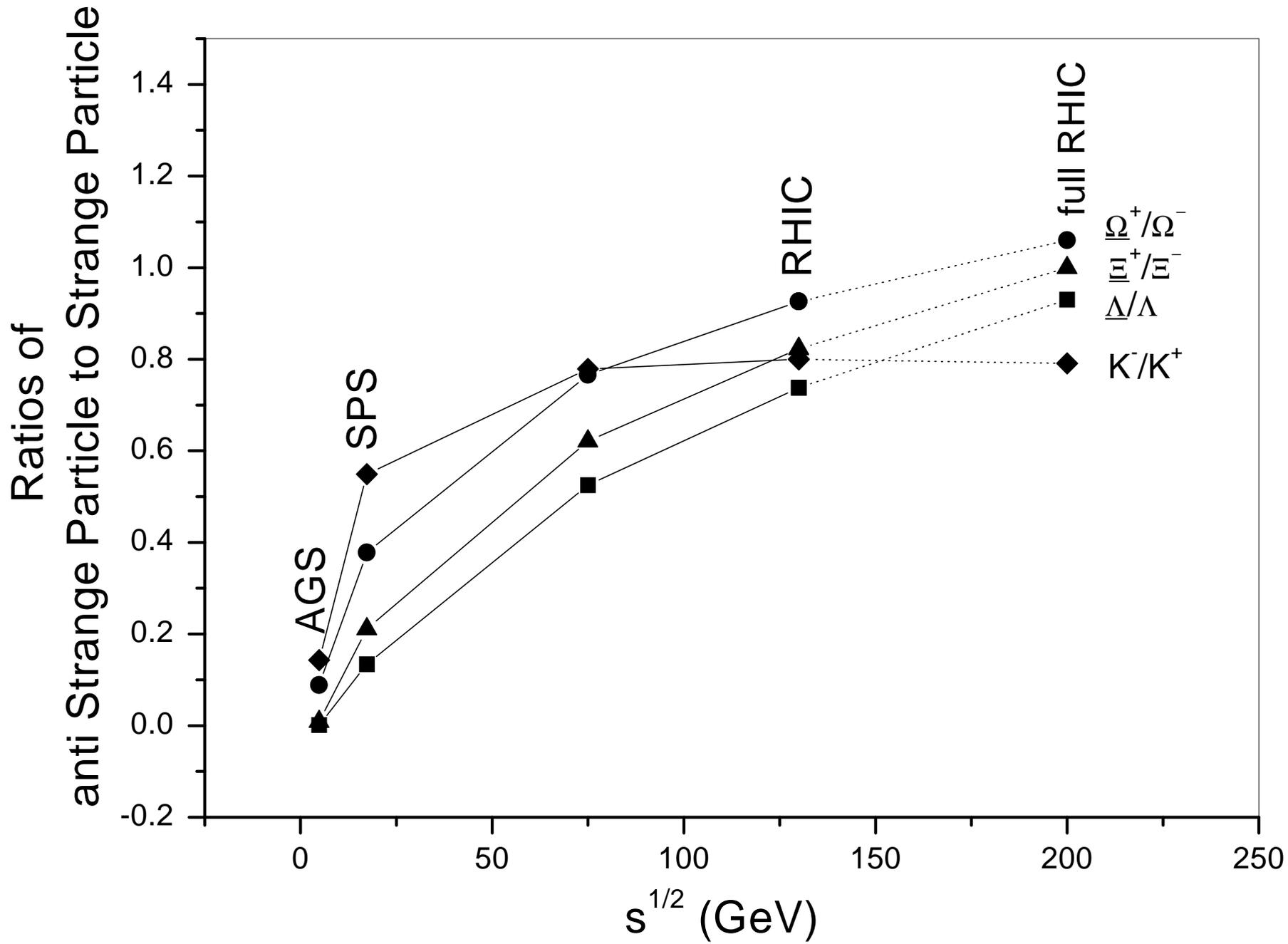
\end{document}